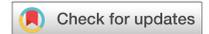

# Low latency FPGA implementation of twisted Edward curve cryptography hardware accelerator over prime field

Md Rownak Hossain[1], Md Sazedur Rahman[2], Kh Shahriya Zaman[3], Walid El Fezzani[4], Mohammad Arif Sobhan Bhuiyan[5✉], Chia Chao Kang[5], Teh Jia Yew[6✉] & Mahdi H. Miraz[6,7,8✉]

The performance of any elliptic curve cryptography hardware accelerator significantly relies on the efficiency of the underlying point multiplication (PM) architecture. This article presents a hardware implementation of field-programmable gate array (FPGA) based modular arithmetic, group operation, and point multiplication unit on the twisted Edwards curve (Edwards25519) over the 256-bit prime field. An original hardware architecture of a unified point operation module in projective coordinates that executes point addition and point doubling within a single module has been developed, taking only 646 clock cycles and ensuring a better security level than conventional approaches. The proposed point multiplication module consumes 1.4 ms time, operating at a maximal clock frequency of 117.8 MHz utilising 164,730 clock cycles having 183.38 kbps throughput on the Xilinx Virtex-5 FPGA platform for 256-bit length of key. The comparative assessment of latency and throughput across various related recent works indicates the effectiveness of our proposed PM architecture. Finally, this high throughput and low latency PM architecture will be a good candidate for rapid data encryption in high-speed wireless communication networks.

The emergence of the modern Internet of things (IoT) infrastructure is a manifestation of the prediction made by Moore's Law (more computation capability due to aggressive technology scaling) and Edholm's Law (more data communication due to modern wireless standards). Nowadays, billions of devices of various cyber-physical systems like smart cities, healthcare, and intelligent transportation are connected through the IoT network. Consequently, the IoT ecosystem has fostered a widespread network of wireless sensor nodes (WSNs) within its application layer, as shown in Fig. 1. Many IoT devices utilises cloud-based data storage solutions due to limited resources, allowing users to access and share data from anywhere over the Internet. However, this approach raises significant security concerns as attackers can manipulate data through unregistered devices deployed within the IoT ecosystem. Therefore, the futuristic IoT ecosystem demands a secure data transmission network to avert unauthorised access and protect sensitive data[1–3]. Asymmetric cryptography or public-key cryptography (PKC) based security protocols can be a viable solution to deal with the privacy aspects of such networks as it avoids key distribution compared to conventional symmetric cryptography[4]. There are several popular methods of PKCs, such as RSA[5], ECC[6,7] as well as Edward curve cryptography (EdCC)[8]. However, ECC has emerged as an intriguing replacement for traditional RSA encryption, owing to its superior strength-per-bit in achieving equivalent levels of security. Hence, ECC can be deployed in the limited-resource IoT environment to achieve fast computation while upholding the intended level of security.

[1]Electrical and Electronic Engineering, Khulna University of Engineering & Technology, Khulna 9203, Bangladesh. [2]Electrical and Electronic Engineering, Hajee Mohammad Danesh Science and Technology University, Dinajpur, Bangladesh. [3]Electrical and Electronic Engineering, Independent University Bangladesh, Dhaka 1229, Bangladesh. [4]Electrical and Electronic Engineering Department, College of Engineering, Gulf University, Sanad 26489, Kingdom of Bahrain. [5]School of Artificial Intelligence and Robotics, Xiamen University Malaysia, 43900 Sepang, Selangor, Malaysia. [6]School of Computing and Data Science, Xiamen University Malaysia, 43900 Sepang, Selangor, Malaysia. [7]School of Computing, Faculty of Arts, Science and Technology, Wrexham University, Wrexham, UK. [8]Faculty of Computing, Engineering and Science, University of South Wales, Swansea, UK. ✉email: arifsobhan.bhuiyan@xmu.edu.my; jiayew.teh@xmu.edu.my; m.miraz@ieee.org

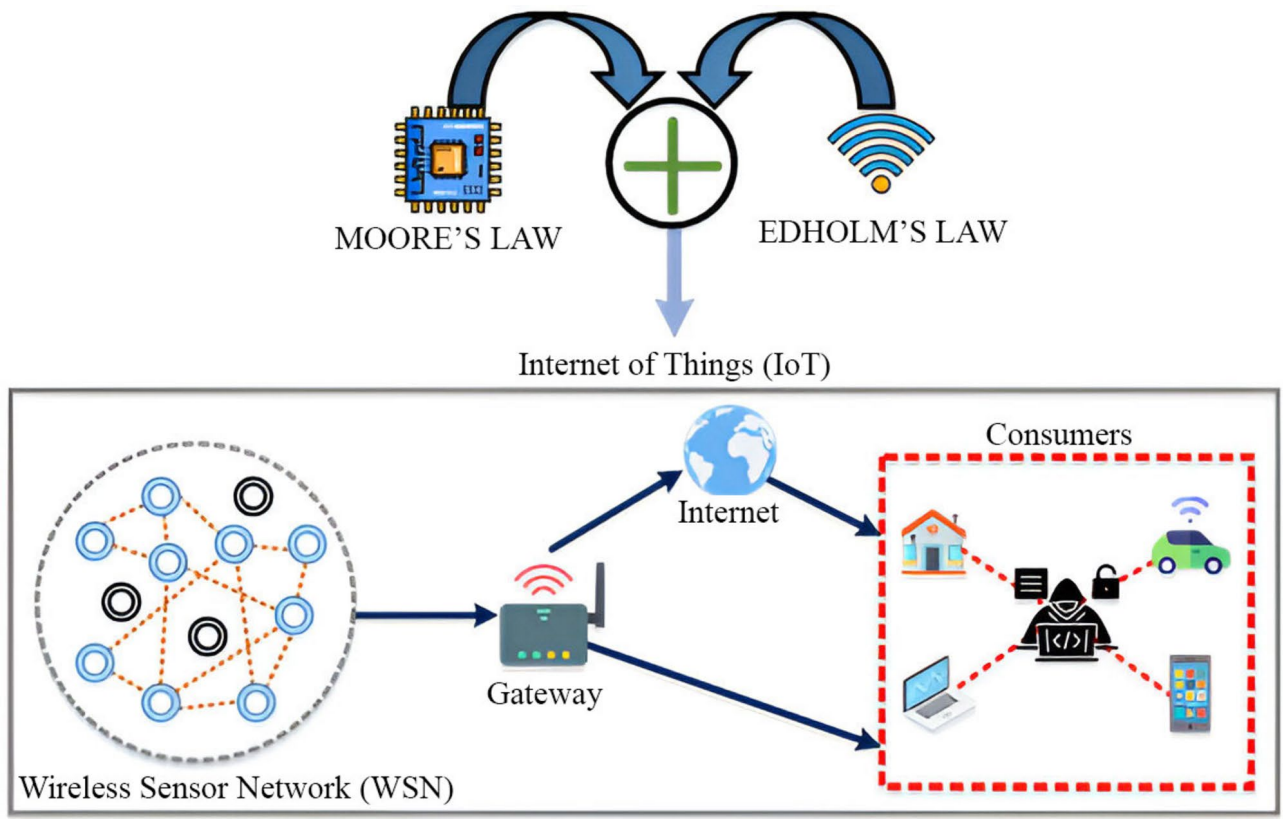

**Fig. 1**. Privacy Aspects of the growing IoT environment.

Side Channel Attacks (SCA) exploit information leaked during the physical implementation of cryptographic algorithms. These attacks target the hardware rather than the cryptographic algorithm itself, making them a significant concern for cryptographic implementations. Power analysis attacks are a type of SCA that involves measuring the power consumption of a cryptographic device during its operation. Simple Power Analysis (SPA) and Differential Power Analysis (DPA) are the two main types of power analysis attacks. SPA examines the power consumption patterns of a device to extract cryptographic keys and other sensitive information. SPA relies on identifying distinct power consumption patterns that correspond to specific operations within the cryptographic algorithm. DPA is a more sophisticated attack that involves statistical analysis of power consumption data collected from multiple cryptographic operations. By analysing the differences in power consumption, attackers can reveal secret information, such as cryptographic keys. Various techniques have been proposed to protect cryptographic hardware implementations against SPA and DPA attacks. For example, the paper by Joye and Tymen[9] discusses countermeasures for ECC against power analysis attacks. Additionally, the study by Sasdrich and Güneysu[10] presents hardware implementations of ECC with protection against SPA and DPA. Unlike Post-Quantum Cryptography[10], which is designed to be secure against attacks from both quantum and classical computers, existing cryptographic systems are susceptible to side-channel attacks. While ECC is widely used and integrated into various cryptographic protocols such as Transport Layer Security, it is important to acknowledge that ECC is not resistant to attacks from quantum computers. Post-Quantum Cryptography aims to develop cryptographic algorithms that are secure against quantum adversaries. Nevertheless, ECC remains a well-established and diffused cryptosystem in current applications, and its efficiency and security are critical for many existing infrastructures.

Edward curves, a special species of the family of elliptic curves, have recently attracted significant research focus because of their high side-channel attack resilience, fast group operation and unified addition formulas. The primary operation of the Edward Curve Crypto Processor (EdCCP) is the Edward curve scalar multiplication (EdCSM) or Edward curve point multiplication (EdCPM), which is expressed as S = k.P; while k is a scalar number, P denotes a particular point on the Edward curve—the resultant point S is found by multiplying an Edward curve point P with a scalar value k. The efficient design of the point multiplication unit is mandatory for developing a high-performance EdCCP, where the performance of the particular point or group operation unit and the modular arithmetic unit determines the efficacy of the point multiplication unit. Thus, optimising the designs of these three units establishes a framework for achieving high efficiency in EdCCP[9–12]. The overall approach of the elliptic curve cryptography (ECC) hardware accelerator design is delineated in Fig. 2. The urge to produce a high-performance ECC accelerator has alluded many researchers to design a high-performance point multiplication unit. Owing to the flexible design environment offered by the FPGA platforms, many FPGA-based hardware architectures for ECC point multiplication on both Galois prime field GF(p) as well as

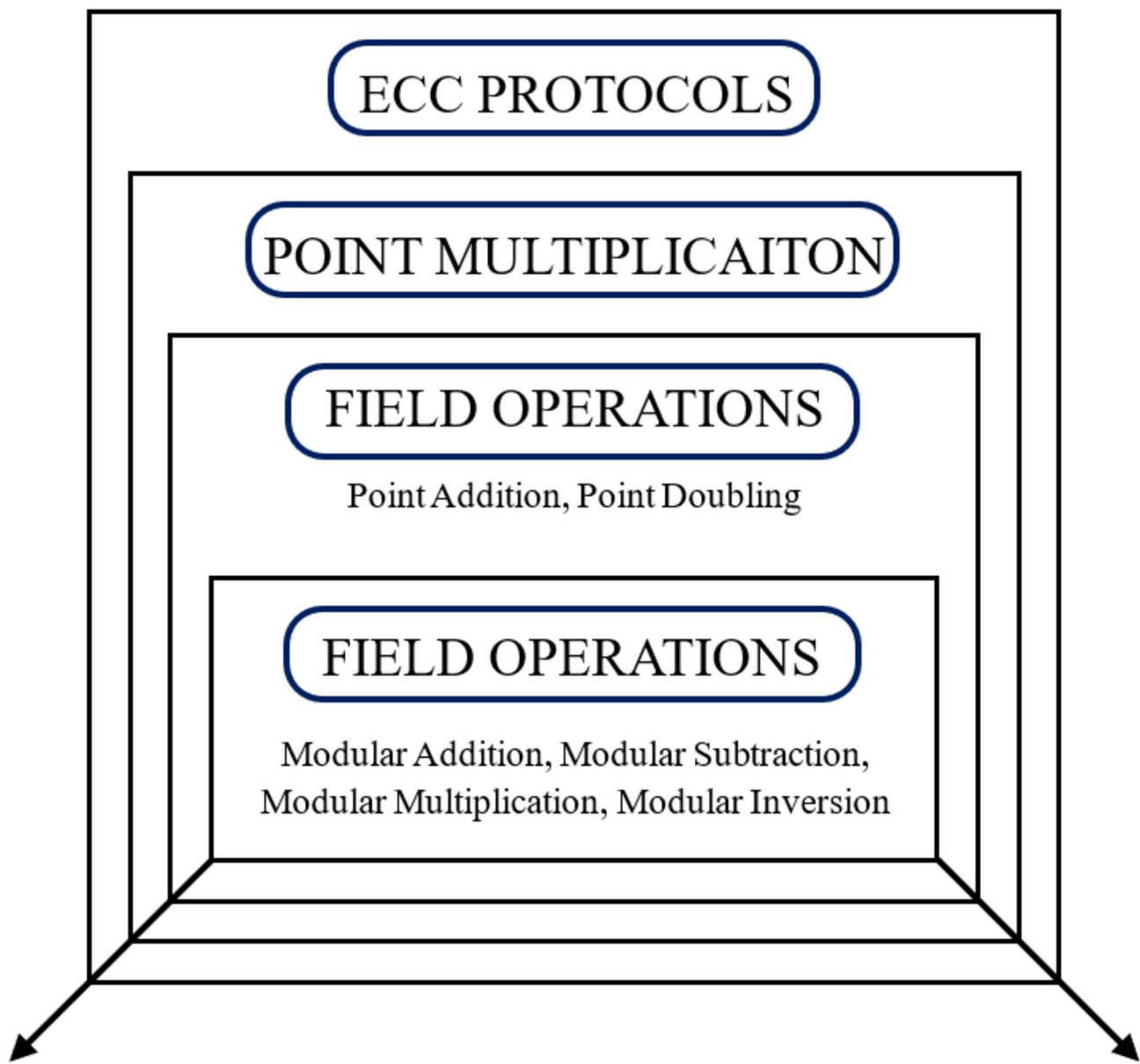

**Fig. 2.** Hierarchy of the ECC hardware accelerator.

Galois binary field GF(2n) have been proposed by many researchers[13–36]. Some of these research were intended to save hardware resources required for small-device applications, while others were intended to minimise the computational time for efficient data encryption in different field sizes. However, a 256-bit architecture over the prime field is preferred most for direct comparison as it is suitable for modern cryptographic security applications. Our proposed EdCPM architecture can be implemented for other standard NIST prime curves.

In Hossain et al.[19] put forward a hardware implementation of the elliptic curve scalar multiplication (ECSM) over a prime field, providing a novel modular multiplication unit using the Montgomery method. In Marzouqi et al.[20] also designed Karatsuba–Ofman modular multiplication unit and Radix-4 Binary GCD Modular Division unit to achieve efficient ECSM operation over NIST curve P256. In the same year, i.e. Amiet et al.[21] also developed a flexible ECSM architecture using an Iterative digit-digit Montgomery algorithm-based modular multiplication unit. The design reported by Salman et al.[22] presents a scalable ECSM unit with mechanisms to prevent side-channel attacks employing Montgomery ladder as well as exponent randomisation to withstand DPA along with SPA. In fact, a cost-effective dual-field ECC processor utilising a word-based Montgomery modular multiplication algorithm was put forward by Lai and Huang[23]. In the period of 2019–2020, Islam et al.[24,25] designed a high-throughput point multiplication module on the twisted Edward curve (Edwards25519) over a 256-bit prime field. In Yeh et al.[26] produced an ECSM unit using a unique technique utilising signed binary representation (SBR) with the M-ary method for reducing the area as well as the energy usage while eschewing SPA. In Lee et al. developed a large field-size ECC processor utilising a novel Montgomery point multiplication (PM) algorithm for minimising the resource consumption while maximising the signal flow[27]. Using the Montgomery ladder approach, Hao et al. devised a lightweight ECSM architecture favouring the random Weierstrass curves over a prime field[29]. In Rashid et al. presented an area-optimised ECC processor for large field-size deploying Lopez-Dahab projective point arithmetic operations[30]. The article by Zhao et al.[31] published in 2021 introduced an ECC processor on the binary field, providing an efficient modular inversion unit using the Itoh–Tsujii inversion algorithm. Between 2021 and 2022, Awaludin et al.[32,35] demonstrated a fast ECSM module using the schoolbook long and Karatsuba multiplication technique for Generic Weierstrass Curves over a prime field. In Kieu-Do-Nguyen et al.[33] described an area-efficient multi-functional ECC processor with a modular inversion unit incorporating the Binary Euclidean algorithm. In Kudithi and Sakthivel[34] implemented an optimised hardware ECC architecture in affine coordinates. In Hu et al.[36] suggested a low-hardware architecture for ECC processors, over GF(p) to be applied in embedded applications, that shows resiliency against SPA attacks.

The primary aim of our design is to achieve an architecture that demonstrates both low latency and high throughput and can be efficiently integrated with current high-speed wireless communication protocols. The major contributions of the research reported in this article, in each unit, are highlighted below:

- A novel EdCPM module on a twisted curve (Edwards25519) has been proposed to accomplish fast computation and high security.
- The EdCPM unit is designed in Jacobian coordinate instead of Affine coordinate to eliminate computationally intensive modular inversion operation.
- An efficient hardware architecture for twisted Edward curve point operation has been designed for minimising arithmetic operations by utilising the parallelisation technique.
- The point operation unit is capable of performing both point doubling (PD) as well as addition (PA) operations within a single operation using a unified point addition formula; thereby offering better resilience against side-channel attacks.
- Multiplication and modular reduction operation are carried out separately utilizing fast reduction modulo as well as Booth radix-4 algorithms to minimise latency and hardware resources.

The subsequent sections of this article are organised along these lines: "Mathematical background" section briefly discusses group operations and field arithmetic on the twisted Edward Curve with relevant equations and algorithms. "Hardware architecture" section outlines proposed hardware architecture for the EdCC accelerator over Edwards25519. Then, "Implementation results" section highlights the implementation results as well as the comparative performance analysis of our EdPM architecture with other existing designs. Lastly, "Conclusions" section summarises and concludes this work.

## Mathematical background
This section presents the mathematical concepts and algorithms associated with the modular arithmetic unit, group operation unit as well as point multiplication unit.

### Finite field arithmetic
The arithmetic of Finite Fields, alias Galois Fields [GF(p)], is a mathematical abstraction of number systems wherein the set of elements in the field (F) is finite. The fundamental operations involved in field arithmetic are Addition and Multiplication. In finite field arithmetic, subtraction operation can be expressed as addition, where $(a, b) \in F$ and $a - b = a + (-b)$. Here, $(-b) \in F$ such that $b + (-b) = 0$. Likewise, inversion/division can be performed in the form of multiplication. However, the inversion unit can be excluded from the Jacobian coordinate system. The finite field's order (q) denotes the elements' number present in any field. As a rule, a finite field is classified as a prime field if its order q could be expressed as a prime power ($q = p^m$), where $m = 1$ and p denotes a prime value[7].

Modular addition as well as subtraction over GF(p) are fundamental cryptosystem operations. Equations (1) and (2) hold the mathematical notation of modular addition and subtraction respectively.

$$Z = (x + y) \bmod p \tag{1}$$

$$Z = (x - y) \bmod p \tag{2}$$

Here, x along with y are the numbers provided, p denotes the prime number and Z signifies the output. The output of the modular addition is derived through the summation of x as well as y $(x + y)$, followed by the deduction of p from $(x + y)$ as long as the resultant (Z) is not less than p. On the other hand, in modular subtraction, if $(x \geq y)$, it could be promptly calculated by simple subtraction or using 2's complement, whereas if $(x < y)$, then y is subtracted from $(x + p)$. However, modular reduction operation is less significant during modular addition and subtraction because the inputs x and y lie from 0 to p – 1, whereby Z must be $\leq 2p$. This paper proposes a combined modular addition and subtraction unit instead of two distinct modules for EdCCP[18].

Modular multiplication is one of the most crucial design units to devise a high-performance cryptosystem, as implementing the modular multiplication over GF(p) requires much area and time compared with other modular arithmetic operations. Generally, the modular multiplication operation can be mathematically expressed by Eq. (3), where M and R are the provided numbers, $p$ denotes the prime number and Z is the output. This research establishes two modules for modular multiplication: one of them for the regular multiplication operation while the other is for the modular reduction operation[18].

$$Z = (M, R) \bmod p \tag{3}$$

### Twisted Edward curve

The mathematical expression of a twisted Edward curve over the field k ($k \neq 2$) is expressed as follows:

$$e_{a,d} : ax^2 + y^2 = 1 + dx^2y^2 \tag{4}$$

Here, $a, d \in GF(p) \setminus \{0,1\}$. In fact, if $\boldsymbol{a} = 1$, it is known as the untwisted Edwards curve. The specifications of the Edwards25519 over GF(p) are: $\boldsymbol{a} = -1$, $d = -121665/121666$ and $\boldsymbol{p} = 2^{255} - 19$[11,12]. Selecting the twisted Edward curve over conventional elliptic curves presents several advantages. Firstly, the twisted Edward curve follows a unified addition law which supports point addition as well as doubling while preserving the identity. Moreover, the Twisted Edward curve saves computational time by offering fewer arithmetic operations than the standard curve[24].

### Projective homogeneous coordinate system

The curve $e_{a,d}$ can be represented within a projective homogeneous coordinate system, where a triplet $(X, Y, Z)$ denotes every point $(x, y)$. This triplet falls in with the affine point $(x = Z/X, y = Z/Y)$, where $Z \neq 0$. Thus, the allied projective twisted Edwards curve can be expressed as:

$$\left(aX^2 + Y^2\right)Z = Z^4 + dX^2Y^2 \tag{5}$$

Several coordinate systems exist, e.g. the projective or Jacobian, affine, Chudnovsky and Lopez-Dahab projective coordinates, for point representation. However, for our research we chose Jacobian coordinates amongst the other popular ones for several reasons. First of all, Jacobian coordinates eliminate the inversion operation which is considered to be the most expensive division, reducing computations on the Edward curve. Secondly, it is possible to present the same affine point (x,y) by Z's various values; hence, such points can be encoded using random values of Z that will offer an extra layer of security against side-channel attacks[28].

### Group law for twisted Edward curve

In twisted Edwards curve, $(0, 1)$ signifies the zero or neutral element while the inverse of any point $(x, y)$ is $(-x, y)$. Let, both $(X_1 : Y_1 : Z_1)$ as well as $(X_2 : Y_2 : Z_2)$ are to be the paired points on the projective twisted Edward curve while $(X_3 : Y_3 : Z_3)$ is the sum of those points[10]. Then, $(X_3 : Y_3 : Z_3)$ can be represented as:

$$X_3 = Z_1Z_2(X_1Y_2 + X_2Y_1)(Z_1^2Z_2^2 - dX_1X_2Y_1Y_2) \tag{6}$$

$$Y_3 = Z_1Z_2(Y_1Y_2 - aX_1X_2)(Z_1^2Z_2^2 + dX_1X_2Y_1Y_2) \tag{7}$$

$$Z_3 = (Z_1^2Z_2^2 - dX_1X_2Y_1Y_2)(Z_1^2Z_2^2 + dX_1X_2Y_1Y_2) \tag{8}$$

### Point multiplication

Computationally intensive point multiplication (PM) is considered to be the most significant function of an EdCC accelerator. Generally, the fundamental process of PM could be characterised as $S = k.P$; while the P denotes a base point within the Edward curve, the k represents a confidential scalar (i.e. the secret/private key) where the S signifies another point within the curve that serves as the public key. Point multiplication can be executed by carrying out an array of point additions and doublings, adopting k's binary bit sequence. The double-and-add method is considered the most forthright approach to execute PM, as outlined in Algorithm 6, wherein point doublings are executed in each cycle. In contrast, point additions are only carried out if $\boldsymbol{k_i} = 1$ [19,24].

### Algorithms

The algorithms used for various mathematical operations, including modular addition, modular reduction, multiplication, subtraction, unified point operation as well as point multiplication, are mentioned below.

```
Input   : Modulus p and integers x, y ∈ [0, p-1]
Output  : Z = (x + y) mod p
1: Set A = x, B = y
2: Set V = A + B
3: if  V ≥ p   then Z = V - p
   else Z = V
4: return Z
```

**Algorithm 1.** Addition in GF(p)[7]

```
Input   : Modulus p and integers x, y ∈ [0, p-1]
Output  : Z = (x - y) mod p
1: Set A = x, B = y
2: if A ≥ B then Z = A − B
   else Z = (A + p) − B
3: return Z
```

**Algorithm 2.** Subtraction in GF(p)[7]

```
Input   : Integers M, R
Output  : Z = M*R
1: Set A = M, B = R, P = Zeros
2: Set S = 2*R
3: for i in 0 to 127 do
       sel = A[i+2]A[i+1]A[i]
     case sel is
          when b "001" | "010"  => P = P+B
          when b "011"          => P = P + 2*B
          when b "100"          => P = P − 2*B
          when b "101" | "110"  => P = P − B
          when others => P = P
     end case
   P = P / 2
4: Z = P
5: return Z
```

**Algorithm 3.** Booth Radix-4 Multiplication[12]

```
Input   : An integer x = (x15,… …, x2,x1,x0) in base
232 with 0 ≤ x <  p256
Output  : x mod p256
1: Define 256-bit integers:
     Sm1 = (x7, x6, x5, x4, x3, x2, x1, x0),
     Sm2 = (x15, x14, x13, x12, x11, 0, 0, 0),
     Sm3 = (0, x15, x14, x13, x12, 0, 0, 0),
     Sm4 = (x15, x14, 0, 0, 0, x10, x9, x8),
     Sm5 = (x8, x13, x15, x14, x13, x11, x10, x9),
     Sm6 = (x10, x8, 0, 0, 0, x13, x12, x11),
     Sm7 = (x11, x9, 0, 0, x15, x14, x13, x12),
     Sm8 = (x12, 0, x10, x9, x8, x15, x14, x13),
     Sm9 = (x13, 0, x11, x10, x9, 0, x15, x14),
2: return (Sm1 + 2Sm2 + 2Sm3 + Sm4 + Sm5 − Sm6 − Sm7 −
Sm8 − Sm9  mod  p256)
```

**Algorithm 4.** Fast Reduction Modulo $p_{256} = 2^{256} - 2^{224} + 2^{192} + 2^{96} - 1$[7,13]

```
Input    : P(X1, Y1, Z1); Q(X2, Y2, Z2) ∈ GF(p)
Output   : (P + Q)(X3, Y3, Z3) ∈ GF(p)

1: A = Z1.Z2 , B = A2 , C1 = a.X1.X2 , C2 = X1.Y2
2: D1 = Y1.Y2 ,  D2 = X2.Y1 , E = d.C2.D2 , F = B − E ,
G = B + E
3: X3 = A.F.(C2 + D2)
4: Y3 = A.G.(D1 − C1)
5: Z3 = F.G
```

**Algorithm 5.** Unified Twisted Edward Curve Point Operation[15]

```
Input   :   Elliptic curve E, an elliptic curve point P
and a scalar d with bits di
Output  : T = d.P
Initialisation: T = P
1:   for i = t-1 downto 0
2:      T = T + T mod n
3:        if di = 1 then T = T + P mod n
4:   return T
```

**Algorithm 6.** Double and Add Algorithm for Point Multiplication[15]

## Hardware architecture

High-performance EdCCP requires efficient designing of modular arithmetic, group operation and point multiplication units. This research proposes five hardware architectures for modeling an EdCCP, which will be elaborated on in this section.

### Modular arithmetic unit

*Combined modular addition-subtraction*

The architecture shown in Fig. 3 starts functioning based on the selected operation (addition or subtraction) that must be carried out. Initially, out of two distinct predetermined values, one will be saved in registers depending on the operation selected. Then, an adder will perform the addition operation and hold the value. This value will be sent to the comparator, which will convert that value into a suitable range based on the selected operation. Finally, the outcome will be a 256-bit value for modular addition or subtraction.

*Booth radix-4 multiplication unit*

The diagram shown in Fig. 4 represents the Booth radix-4 multiplication, which operates following Algorithm 3. Following the reset operation, appropriate values will be stored in prod-reg, state-reg, Q-reg as well as result-reg. Subsequently, the values contained within the result-next, prod-next and state-next shall be modified. The multiplier as well as multiplicand values are be transferred to the prod-reg register as well as mcand-reg, respectively, at the time of the IDLE state. Then, an 8X1 multiplexer determines the proper computation depending on the result-next register's three least significant bits (LSBs); this process will continue throughout the BUSY state. When the final value is reached by counter, a 512-bit value is generated as the multiplication output using 128 Clock Cycles.

*Modular reduction unit*

The proposed hardware architecture presented in Fig. 5 executes modular reduction operation and has been developed using the fast reduction modulo algorithm (refer to Algorithm 4). At the onset of the operation, nine values shall be generated utilising the fast reduction modulo algorithm. After that, all values will undergo processing through left-shifters, not gates and adders to satisfy the necessary addition operation. The outcome of this operation will then be combined with six different pre-defined values. Subsequently, the multiplexers will select the appropriate bits based on the values produced by the adder. Finally, the 256-bit result is achieved from 512-bit input using only one clock cycle.

*Modular multiplication unit*

Figure 6 illustrates the overall methodology underlying our proposed modular multiplication technique. The modular multiplication unit receives two inputs: a 256-bit multiplier as well as a 256-bit multiplicand. Initially, the two 256-bit inputs are processed through a Booth Radix-4 multiplication unit, resulting in a 512-bit output.

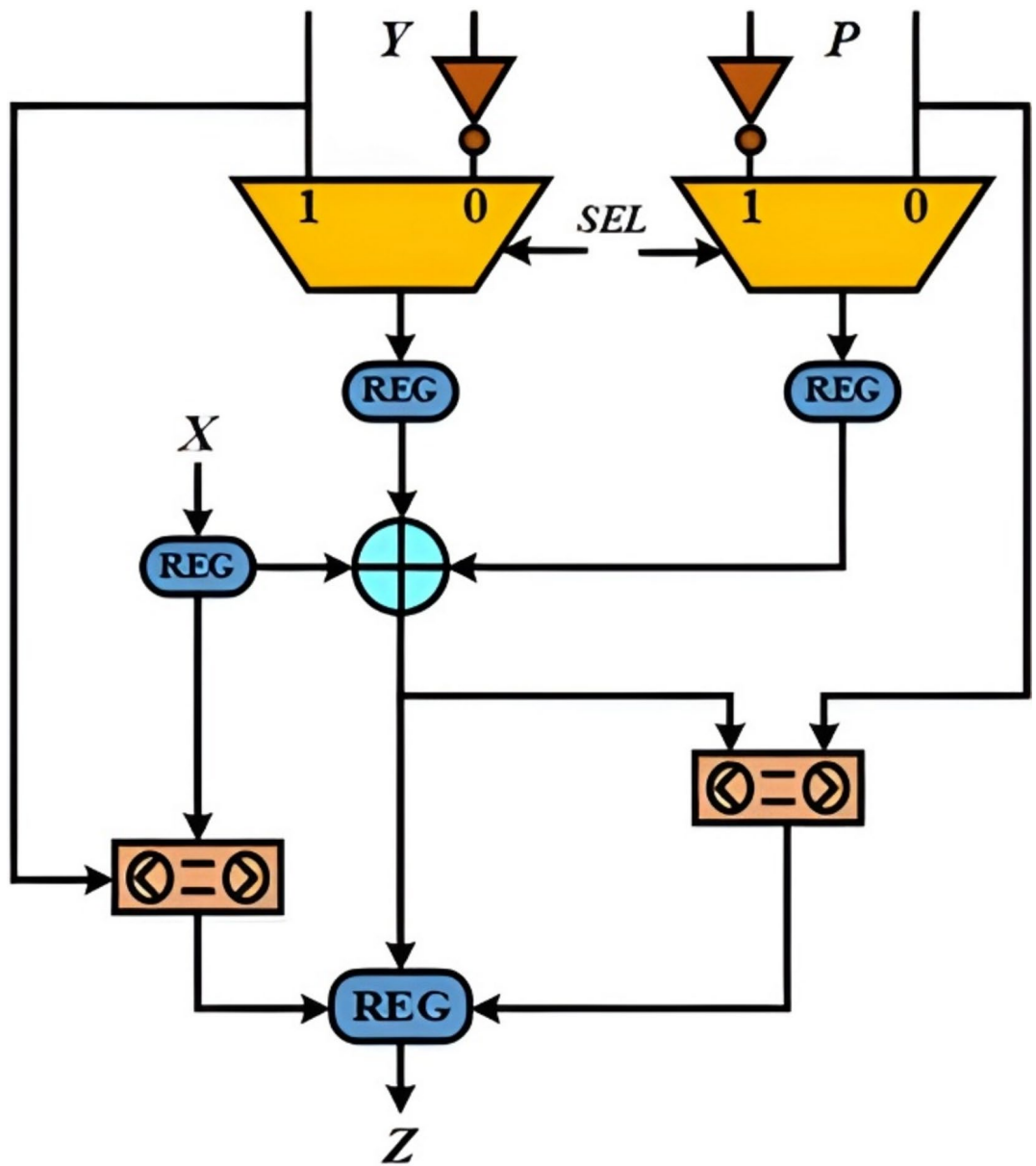

**Fig. 3**. The proposed hardware architecture for combined modular addition as well as subtraction.

Finally, the outcome of the multiplication module, which yields a 512-bit output, undergoes modular reduction architecture to attain a 256-bit output. The complete process of modular multiplication necessitates 129 clock cycles, comprising 128 cycles for multiplication and an additional cycle for modular reduction.

### Group/point operation unit
Elliptic curve group operations comprise modular adders, subtractors, multipliers and squares, distributed across multiple levels to execute point multiplication operations. The group operation module is designed in projective coordinates according to the Unified Point Operation algorithm, as mentioned in Algorithm 5. Figure 7 depicts this unit's hardware design, which has six successive levels that cost thirteen modular multipliers, one modular square operator, two modular additions operators and two modular subtraction operators denoted as (13 M + 1S + 4A). The six distinct levels are shown here to visualise the parallelisation that takes place in the overall operation. In order to reduce arithmetic operations and latency, the proposed group operation architecture is optimised using parallelisation techniques across various levels. In this design, modular multiplication and squaring necessitate $m/2 + 1$ clock cycles, while modular addition as well as subtraction require a single clock cycle to complete the operation. Here, m denotes the total count of bits involved per operation. In addition, computational complexity of a level is determined by squaring as well as multiplication operations. Levels having squaring and multiplication require $m/2 + 1$ clock cycles, while levels without squaring or multiplication require only one cycle to proceed to the following level. Thus, the total clock cycles (CCs) needed for the group operation unit is (5 m/2 + 6) CCs. Thus, for 256-bit, the latency (i.e. CCs) for group operation is 646 CCs.

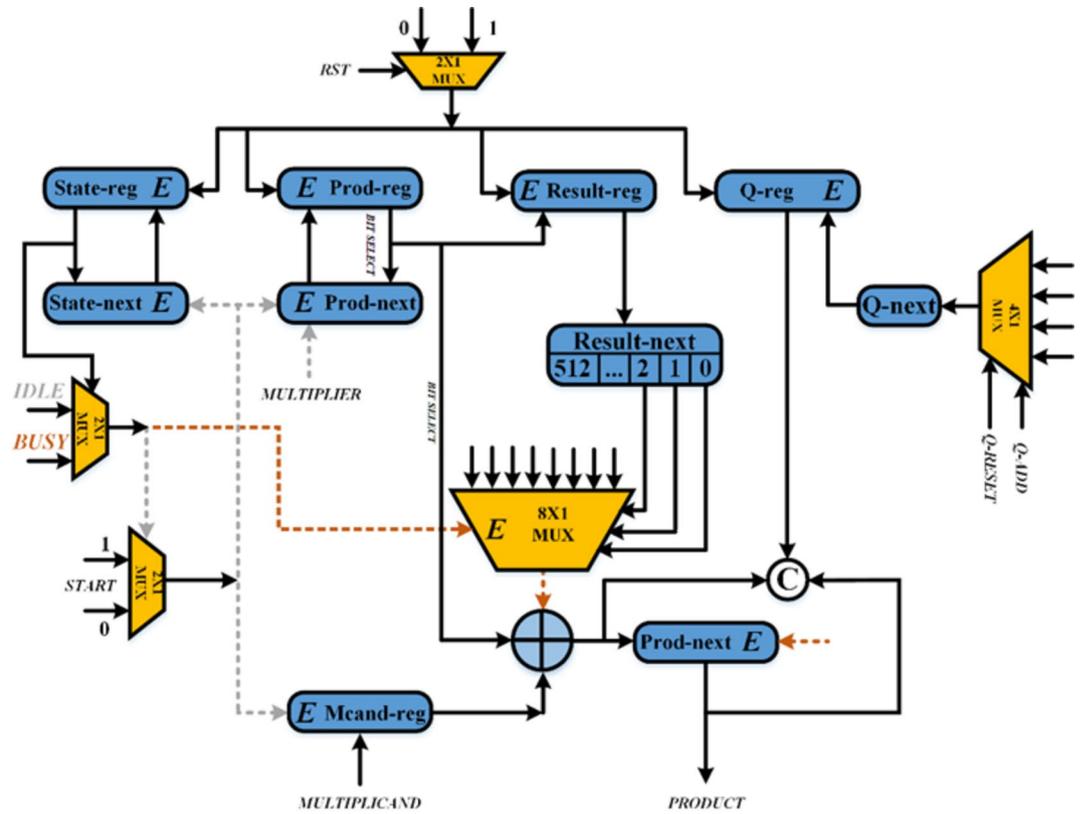

**Fig. 4**. The proposed hardware architecture for booth radix-4 multiplication.

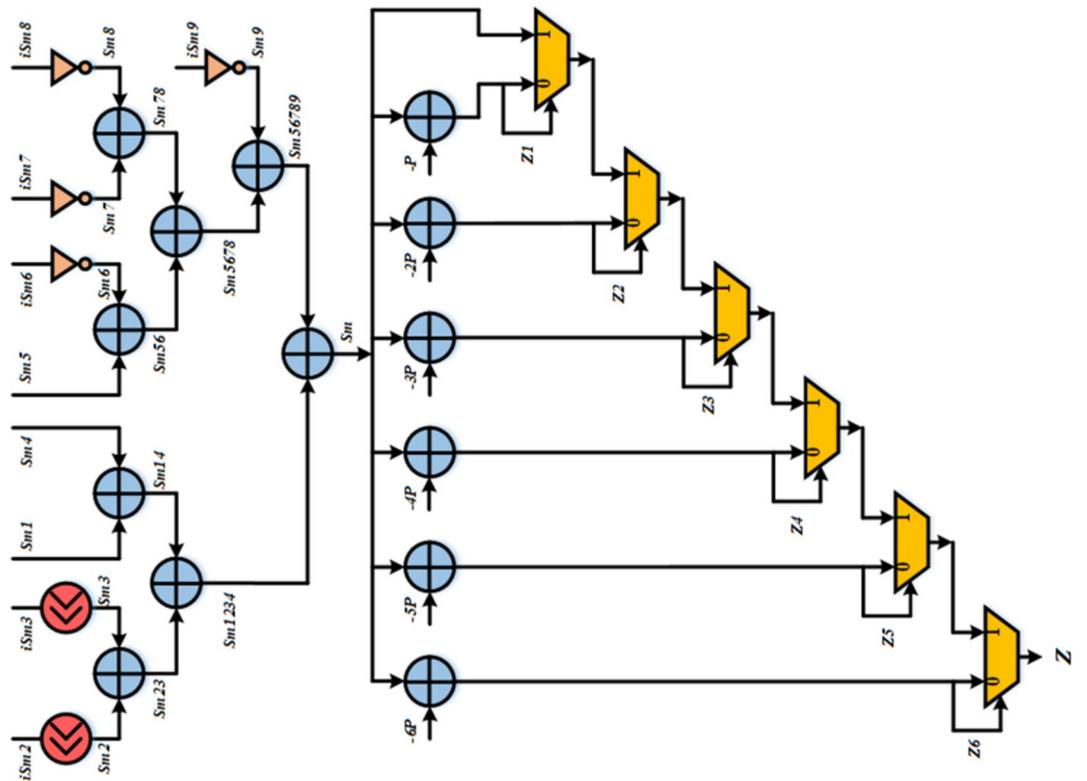

**Fig. 5**. The proposed hardware architecture for modular reduction.

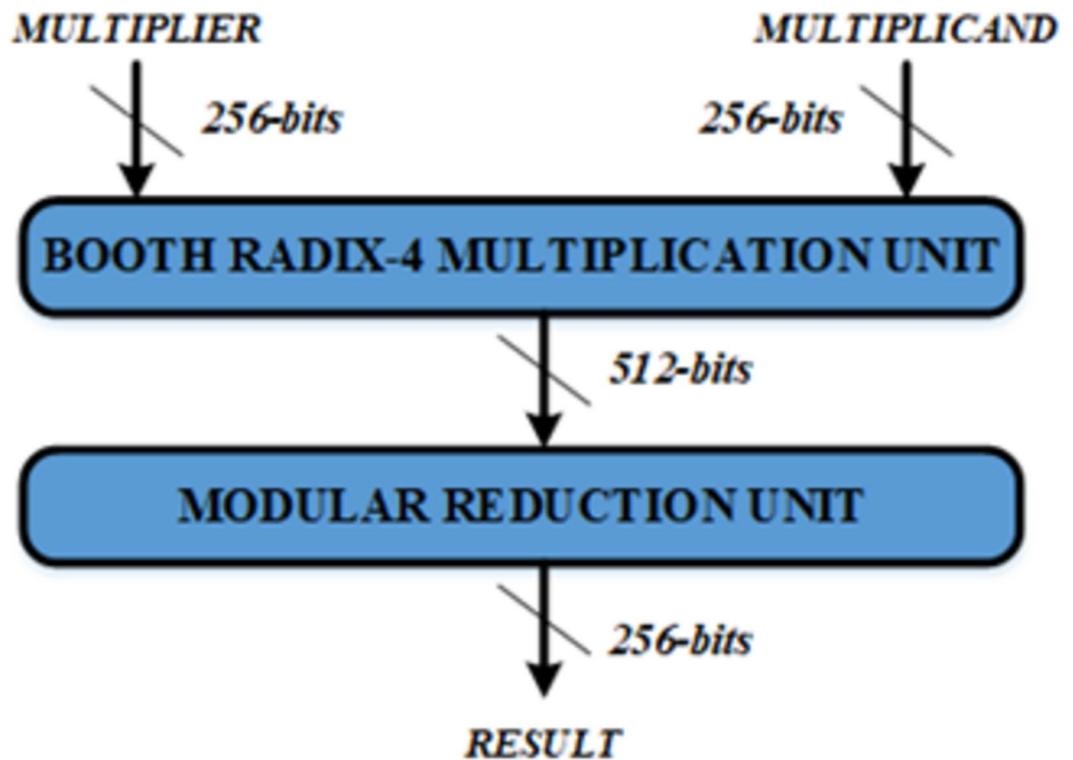

**Fig. 6.** The proffered modular multiplication architecture's block diagram.

#### Point multiplication unit
Figure 8 depicts the proposed EdCPM over the prime field using efficient group operations in Jacobian coordinate. The double and add algorithm is utilised for completion of the proposed point multiplication scheme, as outlined in Algorithm 6. The unified point operation module performs both PD and PA in Jacobian coordinates. The input of PA and the output of PD are compared using the comparative unit. The output of EdCPM is defined as k.P, while k (key) denotes a private key, P denotes a point within the twisted Edward curve. In EdCPM architecture, the input of PD is P (Px, Py, Pz) and the output is Q (Q2x, Q2y, Q2z). The input of PA is P (Px, Py, Pz) + Q (Q2x, Q2y, Q2z) and the output is (Q2px, Q2py, Q2pz). The output of the bit patterns of the input key depends upon the MUX2 output. The total clock cycles required for EdCPM is computed by: CCEdCPM = (m − 1) (CCEdUPO) = (m − 1) (5 m/2 + 6) = (5m2/2 + 7 m/2 − 6). For 256-bit EdCPM, CCEdUPO = 164,730 clock cycles.

#### Implementation results
This section analyses and reports the post-synthesis performance of the preferred modular arithmetic architectures, a point operation unit and a point multiplication unit over GF(p). The proffered EdCC accelerator has been materialised utilising Xilinx ISE 14.5 Design Suite, which was synthesised on the Virtex-5 (xc5vl50t-2ff1136) FPGA platform. The simulations were performed utilising Modelsim and Isim, while the outcomes were verified employing the Maple software. On the Virtex-5 FPGA, the maximum frequency of the proposed modular arithmetic, point operation and point multiplication modules is 117.809 MHz.

Various multiplication architectures have been designed which is depicted in Table 1, among them Booth radix-4 shows the best hardware performance. After that, the best multiplication hardware (Booth radix-4) is selected for modular multiplication with the help of our designed modular reduction module. Based on the implementation results in Table 1, Booth Radix-4 multiplication with the fast reduction modulo is by far the most efficient hardware implementation approach both in terms of optimized area and time having 1290(4%) slices, 4915(17%) LUTs, 584(10%) FFs and 2.04 µs delay. All hardware architectures have been implemented on Virtex-5 FPGA.

A comparative analysis with other relevant works has been presented, in this section, to demonstrate the efficacy of the proposed research. This work employs a unified design approach for modular addition and subtraction instead of discrete execution of these operations to minimise hardware resources. The combined modular addition and subtraction unit operates within a single clock cycle. Thus, this architecture requires only 0.575 ns and 4% of the available slice LUTs for a 256-bit prime field. The proposed modular multiplication architecture is developed by merging the Booth Radix-4 multiplication algorithm as well as the fast reduction modulo algorithm. This modular multiplication approach utilises 1290 slices (constituting 4% of the total), 4915 LUTs (17%), and 584 FFs (10%) and incurs a delay of 2.04 μs. The total count of clock cycles necessary for conducting the multiplication can be determined as (128 + 1) due to the utilisation of the Booth Radix-4 multiplication algorithm, which simultaneously processes two (2) bits and the modular reduction operation, which necessitates one clock cycle. Consequently, the time required for executing the modular multiplication operation calculated to be (15.832 ns × 129), which equals to 2.04 microsecond in Virtex-5 FPGA over GF (256). Furthermore, our proposed architecture for the point operation module on the twisted Edward curve is based on the unified point operation algorithm. The implementation results reveal that it only takes 3102 slices (10% of the available slices) for a prime field of 256-bit in Virtex-5 FPGA. The average time of the point operation unit is (646 X 8.48 ns) = 5.48 μs at 117.809 MHz frequency, where the rate of throughput of this unit is (256/5.48 μs) = 46.72 Mbps. The Edward Curve point multiplication (EdPM) module is designed using the high-performance modular arithmetic and point operation unit that utilises the Double and Add algorithm for optimal efficiency. The EdPM unit exhibits a latency of 164,730 clock cycles, while it requires 1.4 ms to execute single-point multiplication for any 256-bit key with a throughput of 183.38 kbps. Table 2 summarises the implementation outcomes of the proffered EdCCP for a 256-bit prime field.

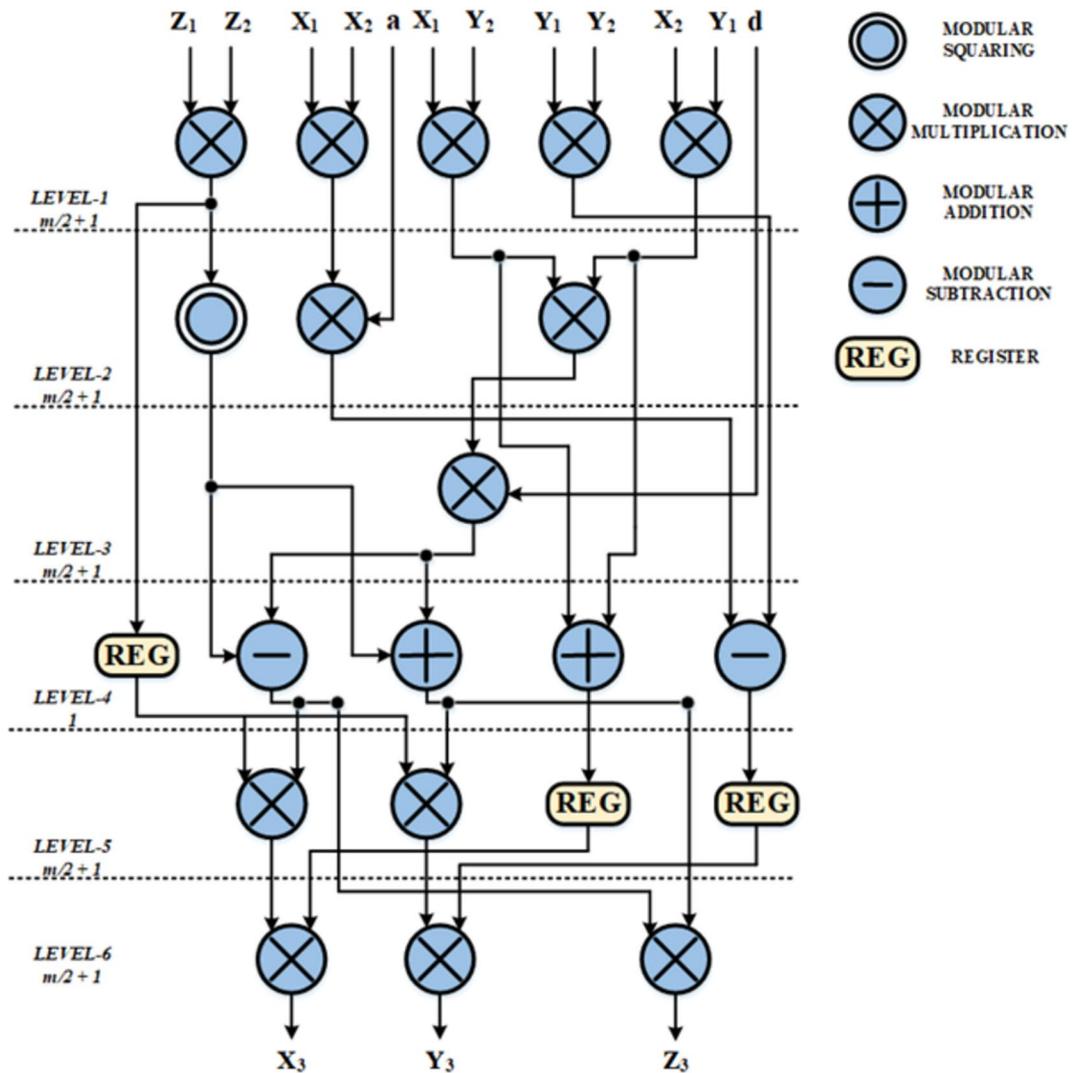

**Fig. 7**. Proposed hardware architecture for unified point operation.

Table 3 presents performance comparisons of our proposed point multiplication (PM) unit and other avant-garde point multiplication designs over GF(p). Hossain et al., 2016 proposed a PM architecture, adopting a Double and Add algorithm that takes 5.26 ms with a corresponding throughput of 48.67 Kbps for executing a point multiplication operation over the prime curve p-256[19]. The proposed accelerator exhibits better speed and throughput compared to the one proposed in[19]. Marzouqi et al.[20] put forward an ECC processor architecture based on RSD that consumes 397,300 CCs to perform a point multiplication, almost 2.5 times greater than our proposed design. Amiet et al.[21] designed a PM architecture using Virtex-7 FPGA platform that completes a point multiplication in 1.49 ms; however, the CCs requirement of this design is higher (335,360). Salman et al.[22] engineered a PM scheme with countermeasures to side-channel attacks where the throughput rate is 34.57 Kbps. Our design offers better latency and side-channel attack resilience due to utilising a unified point operation algorithm than the design recommended in[22]. The dual field PM architecture reported by Lai and Huang[23] necessitates 2.66 ms time which is higher than our design. Our proposed PM architecture shows better design performance with respect to latency compared with that of the other reported designs[24,28,29,33,34]. The processor documented in research by Hu et al.[36] is reconfigurable in terms of various field orders as well as immune to the side-channel attacks. In addition, computational costs of the design in[36] for point multiplication are 610 k clock cycles, whereas our proposed design exhibits lower computational costs of 164.7 k clock cycles for point multiplication. Its ECPM performance requires 29.84 ms, almost 20 times higher than our design. Therefore, our proposed EdCC hardware accelerator will advance the rapid data encryption process especially in high-speed wireless communication networks. The implementations in[37,38] used Intel Agilex, which uses superior technology compared to our Virtex-5. Hence, these implementations can achieve higher clock speeds, lower latency and further acceleration through specialised DSP units. Therefore, our proposed implementation is to yield better results on the Intel Agilex FPGA platform. Besides, Choi et al.

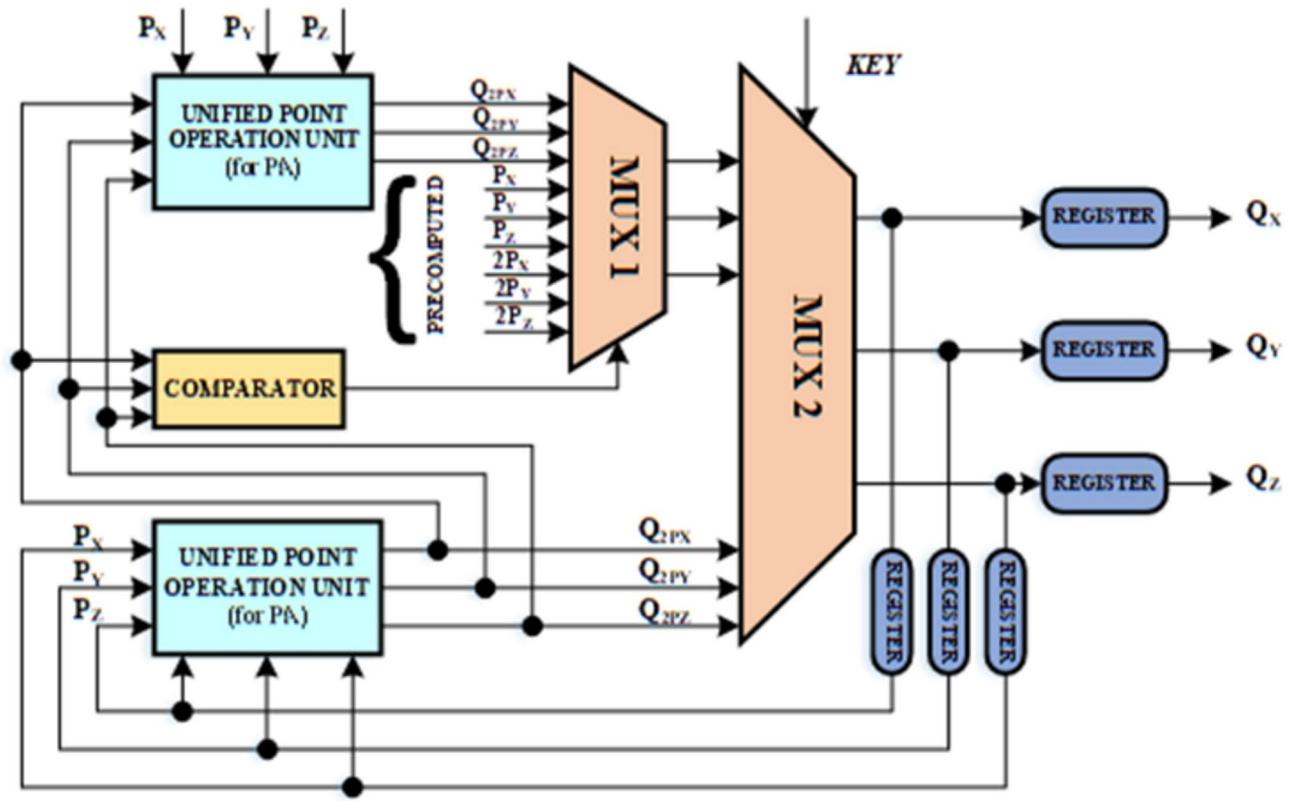

**Fig. 8.** Proposed point multiplication hardware architecture.

| Multiplication approach | Area | | Time (μs) |
|---|---|---|---|
| | Slices | LUTs, FFs | |
| Booth Radix-4 | 1290(4%) | 4915(17%), 584(10%) | 2.04 |
| Booth Radix-2 | 2046(7%) | 4377(15%), 1087(20%) | 4.01 |
| Normal approach | 2055(7%) | 4646(16%), 1033(18%) | 4.07 |
| Moore approach | 1280(4%) | 3357(11%), 363(8%) | 4.01 |

**Table 1.** Overall comparison among our Modular Multiplication Architectures.

| Operation | Platform | Field Size | Clock cycles | Maximum frequency(MHz) | Time | Throughput |
|---|---|---|---|---|---|---|
| Modular Multiplication | Virtex-5 | 256 | 129 | 117.809 | 2.04 μs | 131.9 Mbps |
| point Operation | Virtex-5 | 256 | 646 | 117.809 | 5.48 μs | 46.72 Mbps |
| point Multiplication | Virtex-5 | 256 | 164,730 | 117.809 | 1.4 ms | 183.38 Kbps |

**Table 2.** Results of the implementation of the proposed EdCCp module over GF(256).

| Design | Platform | Frequency (MHz) | Field size | Reported area | Latency (CCs) | Time (ms) | Throughput (Kbps) | Area × Time |
|---|---|---|---|---|---|---|---|---|
| This work | Virtex-5 | 117.809 | 256 | 7.9 K Slices | 164.7 K | 1.40 | 183.38 | 11.06 |
| [19] | Virtex-5 | 75.43 | 256 | 393 K Slices | 397.3 K | 5.26 | 48.67 | 2067.18 |
| [20] | Virtex-5 | 160 | 256 | 34.6 K LUTs | 361.7 K | 2.26 | 113.25 | – |
| [21] | Virtex-7 | 225 | 256 | 6.8 K LUTs + 20 MLUTs | 335.4 K | 1.49 | 171.76 | – |
| [22] | Virtex-7 | 214 | 256 | 1.3 K Slices + 2.7 K LUTs + 4 BRAMS | 1584.9 K | 1.80 | 34.57 | – |
| [23] | Virtex-2 | 94.7 | 256 | 41.6 K Slices | 252.2 K | 2.66 | 96.56 | 110.66 |
| [24] | Virtex-7 | 177.7 | 256 | 8.9 K Slices | 262.7 K | 1.48 | 173.20 | 13.172 |
| [27] | Virtex-7 | 168 | 409 | 11.1 Slices | 21.95 K | – | – | 244 |
| [28] | Virtex-6 | 93.23 | 256 | 6.6 K Slices | 198.6 K | 2.13 | 120.12 | 14.06 |
| [29] | Kintex-7 | 156.3 | 256 | 6.5 K Slices | 270.1 K | 1.73 | 147.9 | 11.25 |
| [33] | Virtex-7 | 149.23 | 256 | 7.2 K Slices | 261.7 K | 1.75 | 146 | 12.6 |
| [34] | Virtex-7 | 124.2 | 224 | 5.4 K Slices | 464.1 K | 3.73 | 68.52 | 20.14 |
| [36] | Virtex-4 | 20.44 | 256 | 6.4 K Slices | 610.0 K | 29.84 | 8.58 | 54.528 |
| [37] | Agilex | 203.96 | 256 | 5.4 K Slices + 15.6 K LUTs + 13.2 K FF + 128 DSP units | 45.3 K | 0.22 | – | – |
| [38] | Agilex | 200 | 256 | – | 5652 | – | – | – |
| [39] | Virtex-6 | 121.6–125.1 | 256 | – | – | 0.30–2.94 | – | 18.6 k – 8.1 k |
| [40] | Virtex-7 | 86.6 | 256 | 12.1 k LUTs | 52.8 K | 0.61 | 420 | 7.4 |
| [41] | Virtex-7 | 120 | 256 | 16,907 LUTs + 4.2 K slices | 23 K | 0.188 | – | – |
| [42] | Virtex-5 | 76.31 | 256 | 8.7 K slices | 300 K | 3.93 | 65.14 | 34.25 |
| [43] | Atrix-7 | 289 | 251 | 6827 slices + 24,778 LUTs | 3263 | 11.29 μs | – | 0.077 |

Table 3. Comparison of the proffered PM unit with other designs over GF(256).

proposed an ECC processor with variable partial product bit[38]. Their FPGA implementation resource usage depends on the selection of partial product bit-width. Both[40,41] presented unconventional architectures, based on residue number system and double-point multiplier respectively, both of which achieved a very high throughput but cost more FPGA resources. In[42], the authors proposed a low-resource using ECC, which traded off substantial throughput. A pipelined approach was proposed in[43], where the authors achieved high performance on their Atrix-7 FPGA with field size of 251. However, the resource usage was significantly higher compared to the other works.

In terms of latency, our point multiplication module requires only 164,730 clock cycles to perform a single-point multiplication, which is significantly lower than many existing designs (e.g.,[19–24,28,29,33,34,36]). As for throughput, our design achieves a throughput of 183.38 kbps, which is higher than most of the compared works (e.g.,[19,20,22–24,28,29,33,34,36]). Considering the area efficiency, our modular multiplication unit uses only 1290 slices (4% of the total available slices) on the Xilinx Virtex-5 FPGA platform, which is highly efficient compared to other designs.

The clock cycles and computation time of our design are quite competitive, ensuring that our design is more efficient for modern high-speed wireless communication standards. Although our designs are implemented in an earlier FPGA technology (Virtex-5), which has higher power consumption and fewer input/output blocks (IoBs), it achieves better outcomes than the other relevant designs.

## Conclusions

Within the scope of this research, a high-speed point multiplication architecture for the EdCC hardware accelerator has been developed using the Edwards25519 curve in a projective coordinate system. An efficient modular multiplier is implemented by adopting Booth Radix-4 Multiplication and Fast modular reduction, which necessitates 129 CCs to multiply two 256-bit integers. A new hardware structure for a group operation unit using a unified point operation algorithm is proposed that requires 646 CCs to execute a single operation. The point multiplication module utilises a double and add always algorithm for faster computation. The designs have been employed on Xilinx Virtex-5 FPGA platform, on a 256-bit prime field. It has been observed that our proposed accelerator completes a point multiplication operation in 164,730 clock cycles, while the processing time is 1.4 ms having a throughput of 183.38 kbps. Our proposed design offers better efficiency in both latency and throughput without compromising security. The comprehensive performance analyses infer that this EdCC will definitely be a viable option for fast and secured data encryption.

## Data availability

All data generated or analysed during this study are included in this published article.

### Acknowledgements
This research was supported by the Xiamen University Malaysia through XMUMRF/2020-C6/IECE/0016 and XMUMRF/2021-C8/IECE/0025 grants as well as the Independent University Bangladesh (IUB) through VCRF-SETS:24-021.

### Author contributions
M.R.H. wrote the main manuscript text, M.S.R. prepared all the figures, M.A.S.B., T.J.Y. and M.H.M. supervised the project, W.E.F. and C.C.K analysed the data, K.H.Z., M.A.S.B. and M.H.M. addressed the review comments. All authors reviewed the manuscript.

### Declarations

### Competing interests
The authors declare no competing interests.

### Additional information
**Correspondence** and requests for materials should be addressed to M.A.S.B., T.J.Y. or M.H.M.

**Reprints and permissions information** is available at www.nature.com/reprints.

**Publisher's note** Springer Nature remains neutral with regard to jurisdictional claims in published maps and institutional affiliations.